\documentclass[10pt]{elsarticle}
\journal{DISCRETE DYN NAT SOC}
\usepackage{amsmath}
\usepackage{indentfirst}
\usepackage{amsfonts,amssymb}
\usepackage{mathrsfs}
\usepackage{amsthm}
\usepackage{lineno,hyperref}
\modulolinenumbers[5]
\newtheorem{Definition}{Definition}[section]
\newtheorem{Theorem}{Theorem}[section]
\newtheorem{Lemma}{Lemma}[section]
\newtheorem{Proposition}{Proposition}[section]
\newtheorem{Example}{Example}[section]
\numberwithin{equation}{section}
\newtheorem{Remark}{Remark}[section]

\begin{document}
	
	\begin{frontmatter}
		
		\title{Regulator-based risk statistics for portfolios}


\author{Xiaochuan Deng\corref{}}
\address{School of Economics and Management, Wuhan University, 	Wuhan 430072,  China}


\ead{dengxiaochuan@whu.edu.cn}
		
	\author{Fei Sun\corref{mycorrespondingauthor}}
\cortext[mycorrespondingauthor]{Corresponding author}
\address{School of Mathematics and Computational Science, Wuyi University, Jiangmen 529020, China}
\ead{fsun.sci@outlook.com}		
		
\begin{abstract}Risk statistic is a critical factor not only for risk analysis but also for financial application. However, the traditional risk statistics may fail to describe the characteristics of regulator-based risk.
	In this paper, we consider the regulator-based risk statistics for portfolios. By further developing the properties related to regulator-based risk statistics, we are able to derive dual representation for such risk. 
\end{abstract}


\end{frontmatter}

	\section{Introduction}

\label{sec:1}
Risk measure is a popular topic in both financial application and theoretical research. 
The quantitative calculation of risk involves
two problems: choosing an appropriate risk model and allocating risk to individual institutions. This has led to further research on risk statistics. 
In a seminal paper,\cite{18} and  \cite{20} first
introduced the class of natural risk statistics with 
representation results. Furthermore, \cite{1} derived an alternative proof for the
natural risk statistics.
Later, \cite{25} and \cite{26}  obtained the representation results for convex risk statistics and quasiconvex risk statistics respectively.

However, traditional risk statistics may fail to describe the characteristics of regulator-based risk.
Therefore, the study of regulator-based risk statistics is particularly interesting. On the other hand, in the abovementioned research on risk statistics, the set-valued risk were never be studied. \cite{19} pointed out that a set-valued risk measure is more appropriate than a scalar risk measure especially in the case where several different kinds of currencies are involved when one is determining capital requirements for the portfolio. Indeed, a natural set-valued risk statistic can be considered as an empirical (or a data-based) version of a set-valued risk measure. More recent studies of set-valued risk measures include those of \cite{2}, \cite{11}, \cite{14}, \cite{15}, \cite{16}, \cite{17}, \cite{21},  \cite{22}  and the references therein.

The main focus of this paper is regulator-based risk statistics for portfolios. In this context, both empirical versions and data-based versions of regulator-based risk measures are discussed.
By further developing the properties related to regulator-based risk statistics, we are able to derive their dual representations. Indeed, This class of risk statistics can be considered as an extension of those introduced by \cite{6}, \cite{9} and \cite{23}.

The remainder of this paper is organized as follows. In Sect.~\ref{sec:2}, we briefly introduce some preliminaries. In Sect.~\ref{sec:3}, we state the main results of regulator-based risk statistics, including the dual representations.
Sect.~\ref{sec:4} investigate the data-based versions of regulator-based risk measures.
Finally, in Sect.~\ref{sec:5}, the main proofs in this paper are discussed.

\section{Preliminary information}
\label{sec:2}

In this section, we briefly introduce some preliminaries that are used throughout this paper. Let $d \geq 1$ be a fixed positive integer. The space $\mathbb{R}^{d\times n}$ represents the set of financial risk positions. With positive values of $X\in \mathbb{R}^{d\times n}$ we denote the gains while the negative denote the losses. Let $n_j$ be the sample size of $D =(X_1, \cdots, X_d)$ in the $j^{th}$  scenario, $j=1, \cdots, l.$  Let $n:= n_1 +\cdots+n_l$. More precisely, suppose that the behavior of $D$
is represented by a collection of data $X=(X_1, \cdots, X_d) \in \mathbb{R}^n \times \cdots \times \mathbb{R}^n$, where
$X_i=(X^{i, 1}, \cdots, X^{i, l}) \in \mathbb{R}^n$,
$X^{i, j} =(x^{i, j}_1, \cdots, x^{i, j}_{n_j}) \in \mathbb{R}^{n_j}$ is the data subset that corresponds to the $j^{th}$ scenario with respect to $X_i$.
For each $j=1, \cdots, l$, $h=1, \cdots, n_j$, $X^j_h:=\left(x^{1, j}_h, x^{2, j}_h, \cdots, x^{d, j}_h\right)$  is the data subset that corresponds to the $h^{th}$ observation  of $D$ in the $j^{th}$ scenario.

In this paper, an element $z$ of $\mathbb{R}^d$ is denoted by $ z:=(z_1, \cdots, z_d).$
An element $X$ of $\mathbb{R}^{d \times n}$ is denoted by
$
X:=(X_1, \cdots, X_d):=\Big(x^{1, 1}_1, \cdots, x^{1, 1}_{n_1}, \cdots, x^{1, l}_1, \cdots,  x^{1, l}_{n_l},
\cdots,  x^{d, 1}_1, \cdots x^{d, 1}_{n_1}, \cdots, x^{d, l}_1, \cdots,  x^{d, l}_{n_l}\Big).
$
Let $K$ be a closed convex polyhedral cone of $\mathbb{R}^{d}$ where $K\supseteq \mathbb{R}^{d}_{++}:=\{(x_{1},\ldots,x_{d})\in \mathbb{R}^{d}; x_{i}>0, 1\leq i\leq d\}$ and $K\cap \mathbb{R}^{d}_{-} = \emptyset$ where $\mathbb{R}^{d}_{-}:=\{(x_{1},\ldots,x_{d})\in \mathbb{R}^{d}; x_{i}\leq 0, 1\leq i\leq d\}$.  Let $K^{+}$ be the positive dual cone of $K$, that is $K^{+}:=\{u\in \mathbb{R}^{d}:u^{tr} v\geq0 \textrm{ for any } v\in K\}$, where $u^{tr}$ means the transpose of $u$. For any $X = (X_{1}, \ldots, X_{d}),  Y = (Y_{1}, \ldots, Y_{d})\in \mathbb{R}^{d\times n}$, $X + Y$ stands for $(X_{1}+Y_{1}, \ldots, X_{d}+Y_{d})$ and $aX$ stands for $(aX_{1}, \ldots, aX_{d})$ for $a\in \mathbb{R}$.
Denote
$K1_n:=\{(z_{1} 1_n, z_{2} 1_n, \cdots, z_{d} 1_n): z \in K\}$ and  $z1_n:=\{(z, z, \cdots, z): z \in \mathbb{R}\}\in \mathbb{R}^n$ where $1_n := (1, \cdots, 1) \in \mathbb{R}^n$.
By $(K1_n)^+$ we denote the positive dual cone of $K1_n$ in $\mathbb{R}^{d \times n}$, i.e.
$(K1_n)^+:=\{w \in \mathbb{R}^{d \times n}: w z^{tr} \geq 0
\text{ for any } z \in K\}$. The partial order respect to $K$ is defined as $a\leq_{K} b$, which means $b-a\in K$ where $a,b\in \mathbb{R}^{d}$ and $X\leq_{K1_n} Y$ means $Y-X\in K1_n$ where $X,Y\in \mathbb{R}^{d\times n}$.

Let $M:=\mathbb{R}^{m}\times \{0\}^{d-m}$ be the linear subspace of $\mathbb{R}^{d}$ for $1\leq m\leq d$. The introduction of $M$ was considered by \cite{14} and \cite{19}. Denote $M_{+}:=M\cap \mathbb{R}^{d}_{+}$ where $\mathbb{R}^{d}_{+}:=\{(x_{1},\ldots,x_{d})\in \mathbb{R}^{d}; x^{i}\geq0, 1\leq i\leq d\}$ and $M^{\bot}:=\{0\}^{m}\times {\mathbb{R}}^{d-m}$. Therefore, a regulator can only accept security deposits in the first $m$ reference instruments. Denote $K_{M}:=K\cap M$ by the closed convex polyhedral cone in $M$, $K^{+}_{M}:=\{u\in M:u^{tr}z\geq0 \textrm{ for any } z\in K_{M}\}$ the positive dual cone of $K_{M}$ in $M$, $intK_{M}$ the interior of $K_{M}$ in $M$. We denote $Q^{t}_{M}:=\{A\subset M:A=clco(A+K_{M})\}$ and $Q^{t}_{M^{+}}:=\{A\subset K_{M}:A=clco(A+K_{M})\}$, where the $clco(A)$ represents the closed convex hull of $A$.


By \cite{7}, a set-valued risk statistic is any map $\rho$ 
\[\rho: \mathbb{R}^{d\times n}\rightarrow 2^{M}\]
that can be considered as an empirical (or a data-based) version of a set-valued risk measure. The  axioms related to this set-valued risk statistic are organized as follows,

[A0] Normalization: $K_{M}\subseteq \rho(0)$ and $\rho(0)\cap -intK_{M}=\phi$;\\
\indent [A1] Monotonicity: for any $X$,$Y\in \mathbb{R}^{d\times n}$, $X-Y\in  K1_{n}$ implies that $\rho(X)\supseteq \rho(Y)$;\\
\indent [A2] M-translative invariance: for any $X\in \mathbb{R}^{d\times n}$ and $z\in \mathbb{R}^{d}$, $\rho(X-z1_{n})=\rho(X)+z$;\\
\indent [A3]Convexity: for any $X,Y\in \mathbb{R}^{d\times n}$ and $\lambda\in[0,1]$, $\rho(\lambda(X)+(1-\lambda)Y)\supseteq \lambda\rho(X)+(1-\lambda)\rho(Y)$;\\
\indent [A4]Positive homogeneity: $\rho(\lambda X)=\lambda\rho(X)$ for any $X\in \mathbb{R}^{d\times n}$ and $\lambda> 0$;\\
\indent [A5]Subadditivity: $\rho(X+Y)\supseteq \rho(X)+\rho(Y)$ for any $X,Y\in \mathbb{R}^{d\times n}$.

We end this section with more notations. A function $\rho: \mathbb{R}^{d\times n}\rightarrow 2^{M}$ is said to be proper if $\textrm{dom}\rho:=\{X\in \mathbb{R}^{d\times n}:\rho(X)\neq \emptyset \}\neq \emptyset $ and $\rho(X)\neq M$ for all $X\in \textrm{dom}\rho$. $\rho$ is said to be closed if  $\textrm{graph}\rho$ is a closed set. The properties of the graphs, see \cite{28}, \cite{29}, \cite{30}.

\section{Empirical versions of regulator-based risk measures}
\label{sec:3}

In this section, we state the dual representations of regulator-based risk statistics, which is the empirical versions of regulator-based risk measures. Firstly, for any $X\in \mathbb{R}^{d\times n}$, $X\wedge_{K1_{n}}0$ is defined as follows
\begin{equation}\label{e21}
X\wedge_{K1_{n}}0:=\left\{ \begin{array}{ll}
X, & \textrm{$X\notin K1_{n}$},\\
0, & \textrm{$X\in K1_{n}$}.
\end{array} \right.
\end{equation}
Therefore, the positions that belongs to $K$ regarded as $0$ position. Next, we derive the properties related to regulator-based risk statistics.

\begin{Definition}\label{D1}
	A regulator-based risk statistic is a function $\varrho:\mathbb{R}^{d\times n}$ $\rightarrow$ $Q^{t}_{M^{+}}$ that satisfies the following properties,
	
	[P0] Normalization: $K_{M}\subseteq \varrho(0)$ and $\varrho(0)\cap -intK_{M}=\phi$;\\
	\indent[P1] Cash cover: for any $z\in K_{M}$, $z\in \varrho(-z1_{n})$;\\
	\indent[P2] Monotonicity: for any $X$,$Y\in \mathbb{R}^{d\times n}$, $X-Y\in \mathbb{R}^{d\times n}\cap K1_{n}$ implies that $\varrho(X)\supseteq \varrho(Y)$;\\
	\indent[P3] Regulator-dependence: for any $X\in \mathbb{R}^{d\times n}$, $\varrho(X)=\varrho(X\wedge_{K1_{n}}0)$;\\
	\indent[P4]Convexity: for any $X,Y\in \mathbb{R}^{d\times n}$ and $\lambda\in[0,1]$, $\varrho(\lambda(X)+(1-\lambda)Y)\supseteq \lambda\varrho(X)+(1-\lambda)\varrho(Y)$.
\end{Definition}

\begin{Remark}
	The property of [P1] means any fixed negative risk position $-z$ can be canceled by its positive quality $z$; [P2] says that if $X_{1}$ is bigger than $X_{2}$ for the partial order in $K$, then the $X_{1}$ need less capital requirement than $X_{2}$, so $\varrho(X_{1})$ contain $\varrho(X_{2})$; [P3] means the regulator-based risk statistics start only from the viewpoint of regulators who only care the positions that need to pay capital requirements, while the positions that belong to $K$  regarded as $0$ position.
\end{Remark}

We now construct an example for regulator-based risk statistics.

\begin{Example}
	The coherent risk measure AV@R was studied by \cite{13} in detail. They have given several representations and many properties like law invariance and the Fatou property. \cite{17} first introduced set-valued AV@R, where the representation result is  derived. Moveover, they also proved that it is a set-valued coherent risk measure. We now define the regulator-based average value at risk. For any $X\in \mathbb{R}^{d\times n}$ and $0<\alpha<1$, we define $\varrho(X)$ as
	\begin{eqnarray*}
		\varrho(X)&:=&AV@R^{loss}_{\alpha}(X)\\
		&:=&\inf_{z\in \mathbb{R}^{d}}\Big\{\frac{1}{\alpha}(-(X\wedge_{K1_{n}}0)|_{M}+z)^{+}-z\Big\}+\mathbb{R}^{m}_{+}.
	\end{eqnarray*}
	It is clear that $\varrho$ satisfies the cash cover, monotonicity, regulator dependence properties and convexity, so $\varrho$ is a regulator-based risk statistic. 
\end{Example}

\begin{Definition}
	Let $Y\in \mathbb{R}^{d\times n}$, $u\in M$. Define a function $S_{(Y,u)}(X):\mathbb{R}^{d\times n}$ $\rightarrow$ $2^{M}$ as
	\begin{displaymath}
	S_{(Y,u)}(X):=\{z\in M:X^{tr} Y\leq u^{tr}z\}.
	\end{displaymath}
\end{Definition}

In fact, the $S_{(Y,u)}(X)$ is the support function of $X$. Before we derive the dual representations of regulator-based risk statistics, the Legendre-Fenchel conjugate theory  (\cite{14}) should be recalled.

\begin{Lemma}\label{L1}
	(\cite{14} Theorem 2) Let $R:\mathbb{R}^{d\times n}$ $\rightarrow$ $Q^{t}_{M}$ be a set-valued closed convex function. Then the Legendre-Fenchel conjugate and the biconjugate of $R$ can be defined, respectively, as
	\begin{displaymath}
	-R^{\ast}(Y,u):=cl\bigcup_{X\in \mathbb{R}^{d\times n}}\Big(R(X)+S_{(Y,u)}(-X)\Big),  \qquad  Y\in  \mathbb{R}^{d\times n}, u\in \mathbb{R}^{d};
	\end{displaymath}
	and
	\begin{displaymath}
	R(X)=R^{\ast\ast}(X):=\bigcap_{(Y,u)\in \mathbb{R}^{d\times n}\times K^{+}_{M}\backslash\{0\}}\Big[-R^{\ast}(Y,u)+S_{(Y,u)}(X)\Big],\qquad X\in \mathbb{R}^{d\times n}.
	\end{displaymath}
\end{Lemma}

\begin{Definition}
	(Indicator function)  For any $Z\subseteq \mathbb{R}^{d\times n}$, the $Q^{t}_{M}$-valued indicator function $I_{Z}: \mathbb{R}^{d\times n}\rightarrow Q^{t}_{M}$ is defined as
	\begin{displaymath}
	I_{Z}(X):=\left\{ \begin{array}{ll}
	clK_{M}, & \textrm{$X\in Z$},\\
	\phi, & \textrm{$X\notin Z$}.
	\end{array} \right.
	\end{displaymath}
\end{Definition}

\begin{Remark}
	The conjugate of $Q^{t}_{M}$-valued indicator function $I_{Z}$ is
	\begin{displaymath}
	-(I_{Z})^{\ast}(Y,u):= cl\bigcup_{X\in Z}S_{(Y,u)}(-X).
	\end{displaymath}
\end{Remark}

\begin{Remark}\label{R1}
	It is easy to see that the regulator-based risk statistic $\varrho$ do not have cash additivity, see \cite{14}. However, $\varrho$ has cash sub-additivity introduced by \cite{10} and \cite{24}.
	Indeed, from the Theorem 6.2 of \cite{15}, $\varrho$ satisfies the Fatou property. Then, consider any $X\in \mathbb{R}^{d\times n}$ and $z\in K_{M}$, for any $\varepsilon\in (0,1)$, we have
	\begin{eqnarray*}
		\varrho\Big((1-\varepsilon)X-z1_{n}\Big)&=&\varrho\Big((1-\varepsilon)X+\varepsilon(-\frac{z}{\varepsilon})1_{n}\Big)\\
		&\supseteq&(1-\varepsilon)\varrho(X)+\varepsilon\varrho(-\frac{z}{\varepsilon}1_{n})\\
		&\supseteq&(1-\varepsilon)\varrho(X)+z
	\end{eqnarray*}
	where the last inclusion is due to the property  [P1]. Using the arbitrariness of $\varepsilon$, we have the following lemma.
\end{Remark}

\begin{Lemma}\label{L2}
	Assume that $\varrho$ is a  regulator-based risk statistic. For any $z\in \mathbb{R}^{d}_{+}$, $X\in \mathbb{R}^{d\times n}$,
	\begin{displaymath}
	\varrho(X-z1_{n})\supseteq\varrho(X)+z.
	\end{displaymath}
	which also implies
	\begin{displaymath}
	\varrho(X+z1_{n})\subseteq\varrho(X)-z.
	\end{displaymath}
\end{Lemma}

\begin{Proposition}\label{P1}
	Let
	$\varrho:\mathbb{R}^{d\times n}$ $\rightarrow$ $Q^{t}_{M^{+}}$ be a proper closed convex regulator-based risk statistic with $u \in 
	\Big\{\Big(-\sum\limits_{j=1}^l\sum\limits_{h=1}^{n_j} Y^{1, j}_h,
	\cdots,
	-\sum\limits_{j=1}^l\sum\limits_{h=1}^{n_j} Y^{d, j}_h\Big) +M^{\perp}\Big\}\bigcap     K^{+}_{M}\backslash\{0\}$.
	Then
	\begin{equation}
	-\varrho^{\ast}(Y,u)=\left\{ \begin{array}{ll}
	cl\bigcup\limits_{X\in \mathbb{R}^{d\times n}}S_{(Y,u)}(-X), & \textrm{$Y \in -{\mathbb{R}_{+}^{d\times n}}\cap (K^{+}1_{n})$},\\
	M, & \textrm{elsewhere}.
	\end{array} \right.\\
	\end{equation}
\end{Proposition}

Now, we state the main result of this paper, the dual representations of regulator-based risk statistics.

\begin{Theorem}\label{T1}
	If $\varrho:\mathbb{R}^{d\times n}$ $\rightarrow$ $Q^{t}_{M^{+}}$ is a proper closed convex regulator-based risk statistic, then
	there is a $-\alpha: (-{\mathbb{R}_{+}^{d\times n}}\cap K^{+}1_{n}) \times K^{+}_{M}\backslash\{0\}$ $\rightarrow$ $Q^{t}_{M^{+}}$, that is not identically $M$ of the set
	\begin{displaymath}
	\mathcal{W}=\bigg\{(Y,u)\in (-{\mathbb{R}_{+}^{d\times n}}\cap K^{+}1_{n}) \times K^{+}_{M}\backslash\{0\}:u \in 
	\Big(-\sum\limits_{j=1}^l\sum\limits_{h=1}^{n_j} Y^{1, j}_h,
	\cdots,
	-\sum\limits_{j=1}^l\sum\limits_{h=1}^{n_j} Y^{d, j}_h\Big) +M^{\perp}\bigg\},
	\end{displaymath}
	such that for any $X\in \mathbb{R}^{d\times n}$,
	\begin{equation}
	\varrho(X)=\bigcap_{(Y,u)\in \mathcal{W}}\Big\{-\alpha(Y,u)+S_{(Y,u)}\big(X\wedge_{K1_{n}}0\big)\Big\}.
	\end{equation}
\end{Theorem}

\section{Alternative data-based versions of regulator-based risk measures}
\label{sec:4}

In this section, we develop another framework, the data-based versions of regulator-based risk  measures. This framework is a little different from the previous one. However, almost all the arguments are the same as those in the previous section. Therefore, we only state the corresponding notations and results, and omit all the proofs and relevant explanations.

We replace $M $ by $\widetilde{M}\in \mathbb{R}^{d\times n}$ that is a linear subspace of $\mathbb{R}^{d\times n}$. We also replace  $K $ by $\widetilde{K}\in \mathbb{R}^{d\times n}$ that is a is a closed convex polyhedral cone where $\widetilde{K}\supseteq \mathbb{R}^{d\times n}_{++}$.
The partial order respect to $\widetilde{K}$ is defined as $X\leq_{\widetilde{K}} Y$, which means $Y-X\in \widetilde{K}$.  Let $\widetilde{M}_{+}:=\widetilde{M}\cap \mathbb{R}^{d\times n}_{+}$. Denote $\widetilde{K}_{\widetilde{M}}:=\widetilde{K}\cap \widetilde{M}$ by the closed convex polyhedral cone in $\widetilde{M}$, $\widetilde{K}^{+}_{\widetilde{M}}:=\{\widetilde{u}\in M:\widetilde{u}^{tr}\widetilde{z}\geq0 \textrm{ for any } \widetilde{z}\in \widetilde{K}_{\widetilde{M}}\}$ the positive dual cone of $\widetilde{K}_{\widetilde{M}}$ in $\widetilde{M}$, $int\widetilde{K}_{\widetilde{M}}$ the interior of $\widetilde{K}_{\widetilde{M}}$ in $\widetilde{M}$. We denote $Q^{t}_{\widetilde{M}}:=\{\widetilde{A}\subset \widetilde{M}:\widetilde{A}=clco(\widetilde{A}+\widetilde{K}_{\widetilde{M}})\}$ and $Q^{t}_{\widetilde{M}^{+}}:=\{\widetilde{A}\subset \widetilde{K}_{\widetilde{M}}:\widetilde{A}=clco(\widetilde{A}+\widetilde{K}_{\widetilde{M}})\}$.
We still start from the viewpoint of regulators who only care the positions that need to pay capital requirements. Therefore, for any $X\in \mathbb{R}^{d\times n}$, we define $X\wedge_{\widetilde{K}}0$ as
\begin{equation}\label{e41}
X\wedge_{\widetilde{K}}0:=\left\{ \begin{array}{ll}
X, & \textrm{$X\notin \widetilde{K}$},\\
0, & \textrm{$X\in \widetilde{K}$}.
\end{array} \right.
\end{equation}
Then, we state the axioms related to regulator-based risk statistics.

\begin{Definition}\label{D41}
	A  regulator-based risk statistic is a function $\widetilde{\varrho}:\mathbb{R}^{d\times n}$ $\rightarrow$ $Q^{t}_{\widetilde{M}^{+}}$ that satisfies the following properties,
	
	[Q0] Normalization: $\widetilde{K}_{\widetilde{M}}\subseteq \widetilde{\varrho}(0)$ and $\widetilde{\varrho}(0)\cap -int\widetilde{K}_{\widetilde{M}}=\phi$;\\
	\indent [Q1] Cash cover: for any $\widetilde{z}\in \widetilde{K}_{\widetilde{M}}$, $\widetilde{z}\in \widetilde{\varrho}(-\widetilde{z})$;\\
	\indent[Q2] Monotonicity: for any $X_{1}$,$X_{2}\in \mathbb{R}^{d\times n}$, $X_{1}-X_{2}\in \mathbb{R}^{d\times n}\cap \widetilde{K}$ implies that $\widetilde{\varrho}(X_{1})\supseteq \widetilde{\varrho}(X_{2})$;\\
	\indent [Q3] Regulator-dependence: for any $X\in \mathbb{R}^{d\times n}$, $\widetilde{\varrho}(X)=\widetilde{\varrho}(X\wedge_{\widetilde{K}}0)$;\\
	\indent[Q4]Convexity: for any $X,Y\in \mathbb{R}^{d\times n}$,\\ $\lambda\in[0,1]$, $\widetilde{\varrho}(\lambda(X)+(1-\lambda)Y)\supseteq \lambda\widetilde{\varrho}(X)+(1-\lambda)\widetilde{\varrho}(Y)$.
\end{Definition}

We need more notations. Let $Y\in \mathbb{R}^{d\times n}$, $\widetilde{u}\in \widetilde{M}$. Define a function $S_{(Y,\widetilde{u})}(X):\mathbb{R}^{d\times n}$ $\rightarrow$ $2^{\widetilde{M}}$ as
\begin{displaymath}
S_{(Y,\widetilde{u})}(X):=\{\widetilde{z}\in \widetilde{M}:X^{tr} Y\leq \widetilde{u}^{tr}\widetilde{z}\}.
\end{displaymath}
Let $\widetilde{R}:\mathbb{R}^{d\times n}$ $\rightarrow$ $Q^{t}_{\widetilde{M}}$ be a set-valued closed convex function. Then the Legendre-Fenchel conjugate and the biconjugate of $\widetilde{R}$ can be defined, respectively, as
\begin{displaymath}
-\widetilde{R}^{\ast}(Y,u):=cl\bigcup_{X\in \mathbb{R}^{d\times n}}\Big(\widetilde{R}(X)+S_{(Y,\widetilde{u})}(-X)\Big),  \qquad  Y\in  \mathbb{R}^{d\times n}, \widetilde{u}\in \mathbb{R}^{d\times n};
\end{displaymath}
and
\begin{displaymath}
\widetilde{R}(X)=\widetilde{R}^{\ast\ast}(X):=\bigcap_{(Y,\widetilde{u})\in \mathbb{R}^{d\times n}\times \widetilde{K}^{+}_{\widetilde{M}}\backslash\{0\}}\Big[-\widetilde{R}^{\ast}(Y,\widetilde{u})+S_{(Y,\widetilde{u})}(X)\Big],\qquad X\in \mathbb{R}^{d\times n}.
\end{displaymath}
For any $\widetilde{Z}\subseteq \mathbb{R}^{d\times n}$, the $Q^{t}_{\widetilde{M}}$-valued indicator function $I_{\widetilde{Z}}: \mathbb{R}^{d\times n}\rightarrow Q^{t}_{\widetilde{M}}$ is defined as
\begin{displaymath}
I_{\widetilde{Z}}(X):=\left\{ \begin{array}{ll}
cl\widetilde{K}_{\widetilde{M}}, & \textrm{$X\in \widetilde{Z}$},\\
\phi, & \textrm{$X\notin \widetilde{Z}$}.
\end{array} \right.
\end{displaymath}
The conjugate of $Q^{t}_{\widetilde{M}}$-valued indicator function $I_{\widetilde{Z}}$ is
\begin{displaymath}
-(I_{\widetilde{Z}})^{\ast}(Y,\widetilde{u}):= cl\bigcup_{X\in \widetilde{Z}}S_{(Y,\widetilde{u})}(-X).
\end{displaymath}
Assume that $\widetilde{\varrho}$ is a  regulator-based risk statistic. For any $\widetilde{z}\in \mathbb{R}^{d\times n}_{+}$, $X\in \mathbb{R}^{d\times n}$,
\begin{displaymath}
\widetilde{\varrho}(X-\widetilde{z})\supseteq\widetilde{\varrho}(X)+\widetilde{z}
\end{displaymath}
which also implies
\begin{displaymath}
\widetilde{\varrho}(X+\widetilde{z})\subseteq\widetilde{\varrho}(X)-\widetilde{z}.
\end{displaymath}

Next, we state the dual representations of regulator-based risk statistics.

\begin{Proposition}
	Let
	$\widetilde{\varrho}:\mathbb{R}^{d\times n}$ $\rightarrow$ $Q^{t}_{\widetilde{M}^{+}}$ be a proper closed convex regulator-based risk statistic with $\widetilde{u} \in 
	\Big\{\Big(-\sum\limits_{j=1}^l\sum\limits_{h=1}^{n_j} Y^{1, j}_h,
	\cdots,
	-\sum\limits_{j=1}^l\sum\limits_{h=1}^{n_j} Y^{d, j}_h\Big) +\widetilde{M}^{\perp}\Big\}\bigcap     \widetilde{K}^{+}_{\widetilde{M}}\backslash\{0\}$.
	Then
	\begin{equation}
	-\widetilde{\varrho}^{\ast}(Y,\widetilde{u})=\left\{ \begin{array}{ll}
	cl\bigcup\limits_{X\in \mathbb{R}^{d\times n}}S_{(Y,\widetilde{u})}(-X), & \textrm{$Y \in -{\mathbb{R}_{+}^{d\times n}}\cap (\widetilde{K}^{+})$},\\
	\widetilde{M}, & \textrm{elsewhere}.
	\end{array} \right.\\
	\end{equation}
\end{Proposition}

\begin{Theorem}
	If $\widetilde{\varrho}:\mathbb{R}^{d\times n}$ $\rightarrow$ $Q^{t}_{\widetilde{M}^{+}}$ is a proper closed convex regulator-based risk statistic, then
	there is a $-\alpha: (-{\mathbb{R}_{+}^{d\times n}}\cap \widetilde{K}^{+}) \times \widetilde{K}^{+}_{\widetilde{M}}\backslash\{0\}$ $\rightarrow$ $Q^{t}_{\widetilde{M}^{+}}$, that is not identically $\widetilde{M}$ of the set
	\begin{displaymath}
	\widetilde{\mathcal{W}}=\bigg\{(Y,\widetilde{u})\in (-{\mathbb{R}_{+}^{d\times n}}\cap \widetilde{K}^{+}) \times \widetilde{K}^{+}_{\widetilde{M}}\backslash\{0\}:\widetilde{u} \in 
	\Big(-\sum\limits_{j=1}^l\sum\limits_{h=1}^{n_j} Y^{1, j}_h,
	\cdots,
	-\sum\limits_{j=1}^l\sum\limits_{h=1}^{n_j} Y^{d, j}_h\Big) +\widetilde{M}^{\perp}\bigg\},
	\end{displaymath}
	such that for any $X\in \mathbb{R}^{d\times n}$,
	\begin{equation}
	\widetilde{\varrho}(X)=\bigcap_{(Y,\widetilde{u})\in \mathcal{W}}\Big\{-\alpha(Y,\widetilde{u})+S_{(Y,\widetilde{u})}\big(X\wedge_{\widetilde{K}}0\big)\Big\}.
	\end{equation}
\end{Theorem}

\section{Proofs of main results}
\label{sec:5}

\noindent \textbf{Proof of Lemma~\ref{L2}.  } The proof of Lemma~\ref{L2} is straightforward from Remark~\ref{R1}.\qed\\

\noindent \textbf{Proof of Proposition~\ref{P1}.}
If $Y\notin -\mathbb{R}_{+}^{d\times n}\cap (K^{+}1_{n})$, there exit an $\bar{X}\in \mathbb{R}^{d\times n}\cap (K1_{n})$ such that $\bar{X}^{tr} Y>0$.
Using the definition of $S_{(Y,u)}$, we have $S_{(Y,u)}(-t\bar{X})=\{z\in M:-t\bar{X}^{tr} Y\leq u^{tr}z\}$ for $t>0$.
Therefore,
\begin{displaymath}
cl\bigcup_{X\in \mathbb{R}^{d\times n}}S_{(Y,u)}(-X)\supseteq \bigcup_{t>0}S_{(Y,u)}(-t\bar{X})=M.
\end{displaymath}
The last equality is due to $-t\bar{X}^{tr} Y\to -\infty$ when $t\to +\infty$.
Using the definition of $S_{(Y,u)}$, we conclude that $cl\bigcup\limits_{X\in \mathbb{R}^{d\times n}}S_{(Y,u)}(-X)\subseteq M$.
Hence
\begin{displaymath}
cl\bigcup_{X\in \mathbb{R}^{d\times n}}S_{(Y,u)}(-X)=M \qquad \textrm{whenever}\qquad Y\notin -\mathbb{R}^{d\times n}\cap (K^{+}1_{n}).
\end{displaymath}
It is easy to check that for any $X\in \mathbb{R}^{d\times n}$ and $v\in M$,
\begin{eqnarray*}
	S_{(Y,u)}(-X-v1_{n})&=&\{z\in M :-X^{tr} Y\leq u^{tr} z+Y^{tr} (v1_{n})\} \\&=&\{z-v\in M:-X^{tr} Y\leq u^{tr} (z-v)+(Y+u1_{n})^{tr} (v1_{n})\}+v\\
	&=&\{z\in M :-X^{tr} Y\leq u^{tr} z+(Y+u1_{n})^{tr} (v1_{n})\}+v.
\end{eqnarray*}
When $(-\sum\limits_{j=1}^l\sum\limits_{h=1}^{n_j} Y^{1, j}_h,
\cdots,
-\sum\limits_{j=1}^l\sum\limits_{h=1}^{n_j} Y^{d, j}_h)+u\in M^\perp$, we have $S_{(Y,u)}(-X-v1_{n})=S_{(Y,u)}(-X)+v$.
However, when $u\notin \big((-\sum\limits_{j=1}^l\sum\limits_{h=1}^{n_j} Y^{1, j}_h,
\cdots,
-\sum\limits_{j=1}^l\sum\limits_{h=1}^{n_j} Y^{d, j}_h)+M^{\perp}\big)$. Therefore,  $(-\sum\limits_{j=1}^l\sum\limits_{h=1}^{n_j} Y^{1, j}_h,\\
\cdots,
-\sum\limits_{j=1}^l\sum\limits_{h=1}^{n_j} Y^{d, j}_h)+u\notin M^\perp$,
we can find $v\in M$ , such that for any $z\in M$,
\begin{displaymath}
-X^{tr} Y\leq u^{tr} z+(Y+u1_{n})^{tr} (v1_{n}).
\end{displaymath}
Therefore, we have
\begin{displaymath}
z+v\in S_{(Y,u)}(-X-v1_{n}).
\end{displaymath}
Therefore
\begin{displaymath}
\bigcup_{z,v\in M}(z+v)\subset \bigcup_{v\in M}S_{(Y,u)}(-X-v1_{n}).
\end{displaymath}
Therefore,
\begin{displaymath}
M\subset \bigcup_{v\in M}S_{(Y,u)}(-X-v1_{n}).
\end{displaymath}
From the definition of $S_{(Y,u)}$, the inverse inclusion is always true. So we conclude that
\begin{displaymath}
M=\bigcup_{v\in M}S_{(Y,u)}(-X-v1_{n}).
\end{displaymath}
It is also easy to check that
\begin{eqnarray*}
	-\varrho^{\ast}(Y,u)&=&cl\bigcup_{X\in \mathbb{R}^{d\times n},v\in M}\Big(\varrho(X+v1_{n})+S_{(Y,u)}(-X-v1_{n})\Big)\\
	&=&cl\bigcup_{X\in \mathbb{R}^{d\times n},v\in M}\Big(\varrho(X+v1_{n})+M\Big)\\
	&=&M
\end{eqnarray*}
where the last equality comes from that the $M$ is a linear space and $\varrho(X)\subseteq M$.
We now derive that $-\varrho^{\ast}(Y,u)=cl\bigcup\limits_{X\in \mathbb{R}^{d\times n}}S_{(Y,u)}(-X)$. In this context,
from $-\varrho^{\ast}(Y,u)=cl\bigcup\limits_{X\in \mathbb{R}^{d\times n}}\Big(\varrho(X)+S_{(Y,u)}(-X)\Big)$,
we derive it in two cases:\\
Case \ 1.\ \ When $X\wedge_{K1_{n}}0=0$, using the definition, we have $\varrho(X)=\varrho(0)\ni0$. Hence
\begin{displaymath}
cl\bigcup_{X\in \mathbb{R}^{d\times n}}\Big(\varrho(X)+S_{(Y,u)}(-X)\Big)\supset cl\bigcup_{X\in \mathbb{R}^{d\times n}}S_{(Y,u)}(-X).
\end{displaymath}
Case\ 2.\ \ When $X\wedge_{K1_{n}}0= X$, we can always find an $\alpha\in K_{M}$ such that $\alpha\in \varrho(X)$. Then
\begin{displaymath}
\varrho(X)+S_{(Y,u)}(-X)\supseteq \alpha+S_{(Y,u)}(-X)=S_{(Y,u)}(-X-\alpha1_{n})=S_{(Y,u)}(-\beta)
\end{displaymath}
where $\beta=X+\alpha1_{n}$. It is relatively simple to check that $\beta\in \mathbb{R}^{d\times n}$. Therefore
\begin{displaymath}
cl\bigcup_{X\in \mathbb{R}^{d\times n}}\Big(\varrho(X)+S_{(Y,u)}(-X)\Big)\supseteq cl\bigcup_{z\in \mathbb{R}^{d\times n}}S_{(Y,u)}(-z),
\end{displaymath}
that is
\begin{displaymath}
-\varrho^{\ast}(Y,u)\supseteq cl\bigcup_{X\in \mathbb{R}^{d\times n}}S_{(Y,u)}(-X).
\end{displaymath}
Consequently, we have
\begin{displaymath}
-\varrho^{\ast}(Y,u)\supseteq cl\bigcup_{X\in \mathbb{R}^{d\times n}}S_{(Y,u)}(-X).
\end{displaymath}
We now need only to derive that $-\varrho^{\ast}(Y,u)\subseteq cl\bigcup_{X\in \mathbb{R}^{d\times n}}S_{(Y,u)}(-X)$.
In fact, for any $z\in \varrho(X)$ and $X\in \mathbb{R}^{d\times n}$, $X+z1_{n}\in \mathbb{R}^{d\times n}$. Therefore
\begin{displaymath}
cl\bigcup_{X\in \mathbb{R}^{d\times n}}S_{(Y,u)}(-X)=cl\bigcup_{X\in \mathbb{R}^{d\times n}}S_{(Y,u)}(-X)\supseteq S_{(Y,u)}(-X-z1_{n})=S_{(Y,u)}(-X)+z.
\end{displaymath}
Using the arbitrariness of $z$, we have
\begin{displaymath}
\varrho(X)+S_{(Y,u)}(-X)\subseteq cl\bigcup_{X\in \mathbb{R}^{d\times n}}S_{(Y,u)}(-X).
\end{displaymath}
Therefore,
\begin{displaymath}
-\varrho^{\ast}(Y,u)\subseteq cl\bigcup_{X\in \mathbb{R}^{d\times n}}S_{(Y,u)}(-X).\ \ \  \ \  \ \ \ \ \ \ \ \ \ \ \ \ \ \ \ \ \ \ \ \ \ \ \ \ \ \ \ \ \ \ \ \ \ \ \ \ \ \ \ \ \ \ \ \ \ \ \ \ \ \ \ \ \ \ \ \qed
\end{displaymath}

\noindent \textbf{Proof of Theorem~\ref{T1}.}
The proof is straightforward from Lemma~\ref{L1} and Proposition~\ref{P1}. \qed

\section*{Conflict of Interest Statement}

The authors declare that the research was conducted in the absence of any commercial or financial relationships that could be construed as a potential conflict of interest.\\
\indent This manuscript has been released as a pre-print at arXiv: 1904.08829v4.

\section*{Funding Statement}
 Funds of Education Department of Guangdong (2019KQNCX156).

\section*{Data Availability Statement}

No data, code were generated or used during the study.



\begin{thebibliography}{99}
	{
		\bibitem{1}   Ahmed, S., Filipovi\'{c}, D., and  Svindland, G. (2008). A note on natural risk statistics, Oper. Res. Lett. 36, 662-664.
		\bibitem{2}  Ararat, C.,  Hamel, A.H., and Rudloff, B. (2017). Set-valued shortfall and divergence risk measures, arXiv: 1405.4905v2 [q-fin.RM] 19 May.
		
		
		\bibitem{4}  Ben-Tal, A., Teboulle, M. (2007). An old-new concept of convex risk measures: the optimized certainty equivalent, Math. Finance, 17(3), 449-476.
		\bibitem{6}  Chen, Y.H., Sun, F., and  Hu, Y.J. (2018). Coherent and convex loss-based risk measures for portfolio vectors, Positivity, 22(1), 399-414. 
		\bibitem{7} Chen, Y.H., Hu, Y.J. (2017). Set-valued risk statistics with scenario analysis, Statist. Probab. Lett. 131, 25-37.
		
		
		\bibitem{9}  Cont, R., Deguest, R., and  He, X.D. (2013). Loss-based risk measures, Stat. Risk Model., 30(2), 133-167.
		\bibitem{27}  Deng, X.,  Sun, F. (2020). Regulator-based risk statistics for portfolios, arXiv: 1904.08829v4 [q-fin.RM] 19 May.
		\bibitem{10}  EL Karouii, N.,  Ravanelli, C. (2009). Cash subadditive risk measures and Interest rate ambiguity, Math. Finance, 19, 561-590.
		\bibitem{11}  Farkas, W.,  Koch-Medina, P., and  Munari, C. (2015). Measuring risk with multiple eligible assets, Math. Financ. Econ. 9(1), 3-27.
		
		
	\bibitem{13}  F\"{o}llmer, H.,  Schied,  A. (2011). Stochastic finance: an introduction in discrete time, Walter de Gruyter Berlin New York, third revised and extended edition.
		
		
		\bibitem{14}  Hamel, A.H. (2009). A duality theory for set-valued functions I: Fenchel conjugation theory, Set-Valued Var. Anal. 17(2), 153-182.
		\bibitem{15}  Hamel, A.H., Heyde, F. (2010). Duality for set-valued measures of risk, SIAM J. Financial Math. 1(1), 66-95.
		\bibitem{16}  Hamel, A.H.,  Heyde, F., and Rudloff, B. (2011). Set-valued risk measures for conical market models, Math. Financ. Econ. 5(1), 1-28.
		\bibitem{17} Hamel, A.H.,  Rudloff, B., and  Yankova, M. (2013). Set-valued average value at risk and its computation, Math. Financ. Econ. 7(2), 229-246.
		\bibitem{18}  Heyde, C.C.,  Kou, S.G., and  Peng, X.H. (2007). What is a good external risk measure: Bridging the gaps between robustness, subadditivity, and insurance risk measures, Working paper, Columbia University.
		\bibitem{19}  Jouini, E.,  Meddeb, M., and  Touzi, N. (2004). Vector-valued coherent risk measures, Finance Stoch. 8(4), 531-552.
		\bibitem{20}  Kou, S.G.,  Peng, X.H., and  Heyde, C.C. (2013). External risk measures and basel accords, Math. Oper. Res. 38, 393-417.
		\bibitem{28}  Liu, J., Wang, C.,  Wang, S.,  Wei, B. (2019). Zagreb Indices and Multiplicative Zagreb Indices of Eulerian Graphs, Bull. Malays. Math. Sci. Soc. 42, 67-78.
		\bibitem{29} Liu, J.,  Zhao, J.,  Min, J.,  Cao, J. (2019). On the Hosoya index of graphs formed by a fractal graph, Fractals-Complex Geometry Patterns and Scaling in Nature and Society, 27(8) 19-35.
		\bibitem{30} Liu, J.,  Zhao, J.,   He, H.,  Shao, Z. (2019). Valency-Based Topological Descriptors and Structural Property of the Generalized Sierpinski Networks, J. Stat. Phys. 177, 1131-1147.
		\bibitem{21}  Labuschagne, C.C.A.,  Offwood-Le Roux, T.M. (2014). Representations of set-valued risk measures definded on the $l$-tensor product of Banach lattices, Positivity, 18(3), 619-639.
		\bibitem{22}  Molchanov, I., Cascos, I. (2016). Multivariate risk measures: a constructive approach based on selections, Math. Finance, 26(4), 867-900.
		\bibitem{23}  Sun, F.,  Chen, Y.H., and  Hu, Y.J. (2018). Set-valued loss-based risk measures, Positivity, 22(3), 859-871.
		\bibitem{24} Sun, F.,  Hu, Y.J. (2019). Set-valued cash sub-additive risk measures, Probab. Engrg. Inform. Sci. 33(2), 241-257.
		\bibitem{25}  Tian, D.J.,  Jiang, L. (2015). Quasiconvex risk statistics with scenario analysis, Math. Financ. Econ. 9, 111-121.
		\bibitem{26} Tian, D.J.,  Suo, X.L. (2012). A note on convex risk statisitc, Oper. Res. Lett. 40, 551-553.
		
		
		
	}
\end{thebibliography}
\end{document}